\documentclass[aps,prb,reprint,showpacs,citeautoscript,
               longbibliography,floatfix]{revtex4-1}

\usepackage{amsmath,amssymb,amsfonts,graphicx,longtable}

\newcommand*{\ket}[1]{| #1 \rangle}

\begin{document}

\title{Demonstration of double EIT using coupled harmonic oscillators
       and RLC circuits}
\author{Joshua Harden}
\author{Amitabh Joshi}
\email{ajoshi@eiu.edu}
\affiliation{Department of Physics, Eastern Illinois University,
             Charleston, Illinois 61920}
\author{Juan D. Serna}
\email{serna@uamont.edu}
\affiliation{School of Mathematical and Natural Sciences,
             University of Arkansas at Monticello, Monticello, Arkansas 71656}

\date{December 31, 2010}

\begin{abstract}
Single and double electromagnetically induced transparency (EIT) in a medium,
consisting of four-level atoms in the inverted-Y configuration, are discussed
using mechanical and electrical analogies. A three coupled spring-mass system
subject to damping and driven by an external force is used to represent the
four-level atom mechanically. The equations of motion of this system are solved
analytically, which revealed single and double EIT. On the other hand, three
coupled RLC circuits are used, as the electrical analog, to explore and
experimentally demonstrate single and double EIT. The simplicity of these two
models makes this experiment appropriate for undergraduate students and easy to
incorporate into a college physics laboratory.
\end{abstract}

\pacs{01.50.My, 42.50.Gy, 42.50.Hz}
\keywords{Coupled Harmonic Oscillators, Coupled RLC Circuits, EIT, single EIT,
          double EIT}

\maketitle

\section{Introduction}

Atomic media have the physical characteristic of absorbing light of certain
frequencies. It is possible, for example, that a medium absorbs two slightly
different light frequencies simultaneously. However, and perhaps more
intriguing, we observe that, for certain atomic configurations of the medium,
the level of absorption of one of the frequencies can be controlled by the other
frequency making the medium virtually transparent to the former frequency. This
phenomenon is called electromagnetically induced transparency
(EIT)~\cite{Harris:36,*Fleischhauer:633}. Usually, EIT occurs in vapors of
three-level atomic systems, where laser lights (coherent light sources) drive
two different atomic transitions sharing one common level (known as the
``probe'' and ``coupling'' field transitions). In the same way that a ``coupling
field'' controls the properties of an EIT medium determining the amount of
absorption of a ``probe field'', the dispersive properties of the medium also
get modified resulting in to the reduction of the group velocity of light inside
it. Physically, EIT can be understood as a process of quantum interference
between two atomic states of a medium involving two indistinguishable quantum
paths that lead to a common final state. In addition to the EIT phenomenon,
double EIT occurs when a four-level atomic system is exposed to three laser
sources driving three different transitions with one common level. The three
transitions are described as the ``probe'', the ``coupling'', and the
``pumping'' field transitions. In this case, two strong electromagnetic fields,
i.e., the coupling and the pumping fields control the medium in determining the
absorption and propagation of the probe field.

The phenomenon of EIT, first observed two decades ago using high-power lasers in
strontium vapor~\cite{Boller:66}, has been extensively investigated during the
past years in atomic beams\cite{Firstenberg:77,*Zhang:80},
plasma\cite{Litvak:88,*Shvets:89}, optical
cavities\cite{Werner:61,*Bentley:61,*Dantan:69,*Yang:95}, and Bose-Einstein
condensates~\cite{Kuang:76,*Weatherall:78}. It has also been studied
theoretically and experimentally for media consisting of three- and four-level
atoms~\cite{Yamamoto:58,*Joshi:67,*Brown:70,*Yang:72,*Olson:116,Joshi:370,
*Zhang:99,*Li:101,*Joshi:79}.

Besides absorption of light, there are other substantial changes observed if a
medium exhibits EIT, such as the modified index of refraction~\cite{Xiao:666},
which can give rise to the reduction of the group velocity of a light
pulse~\cite{Hau:397}, or even a complete stop of light in the
medium~\cite{Liu:409}. Important applications of EIT include lasing without
population inversion~\cite{Mompart:R7,*Wu:78}, enhanced nonlinear optical
processes\cite{Wang:87}, quantum computation and
telecommunications\cite{Schmidt:76,*Ottaviani:90}, quantum
memory~\cite{Hetet:77}, and optical switches~\cite{Bermel:74}.

During the past two decades, the study of quantum-classical analogies in physics
has gained some momentum as they prove to be very useful in helping to
understand the fundamental physical concepts and the applicability of different
theories~\cite{Dragoman}. It is important to note that these analogies bring to
light the fact that similar mathematical models can be applied to both quantum
and classical phenomena, though these theories differ both in formalism, and
fundamental concepts. Recently, a number of these classical analogies of
different quantum optical systems have been reported. For example, stimulated
resonance Raman effect\cite{Hemmer:88}, rapid adiabatic passage in atomic
physics~\cite{Shore:77}, vacuum Rabi oscillation~\cite{Zhu:64}, number-phase
Wigner function and its relation to usual Wigner function~\cite{Vaccaro:243},
and EIT in three-level systems\cite{Garrido:37}. In a recent work,  the response
of a coupled array of nonlinear oscillators to parametric excitation is
calculated in the weak nonlinear limit using secular perturbation theory and the
exact results for small arrays of oscillators are used to guide the analysis of
the numerical integration of the model equations of motion for large arrays.
Such results provide qualitative explanations for experiments involving a
parametrically excited micromechanical resonator array\cite{Lifshitz:67}.

Double EIT phenomenon is very important in EIT based atomic memory systems.
Systems displaying multiple EIT could be useful in the bifurcation of quantum
information in multiple channels temporarily, which then can be used in
multiplexing required in certain quantum information protocols. The release of
stored information from multiple channels could be separately controlled by
manipulating the group velocity of individual channels (via their control
fields) in such systems. Hence double EIT is an important phenomenon for quantum
information processing and quantum computing and thus, it needs its introduction
and realization in the simplest form to the readers.

The goal of this work is to demonstrate double EIT in four-level systems using
two classical analogies: mass-spring systems and RLC circuits. For that purpose,
we first describe the atom as a damped, harmonic oscillator driven by an
external force~\cite{Allen-Eberly}. Three different masses connected by springs
and subject to frictional forces (damping) are used to represent the four-level
atom. The destructive interference of the normal modes of oscillation of the
masses is equivalent to the quantum interference that originates EIT. Secondly,
we explore experimentally, the electrical analogue of double EIT using three
coupled RLC circuits. The power delivered to one of these coupled oscillating
circuits is measured as a function of the frequency of a driving source of
alternating voltage. The electrical equivalence of the power transmitted to the
circuit with the power absorbed by an atomic medium, allows us to investigate,
directly from the circuit, the characteristic patterns of single and double EIT.

To get information about the absorption and dispersion of light in the
four-level atomic medium, we need to solve a large system of the density matrix
equations numerically~\cite{Joshi:370}. However, the equations of motion that
describe the mechanical and electrical systems can be solved analytically and
hence the double EIT phenomenon could be studied with more ease in the two
analogue systems using the analytical solutions. The merit of analytic solutions
is that it clearly brings out the functional dependence of double EIT phenomenon
on several parameters. On the other hand, the circuits used in this experiment
show realistic forced, damped harmonic oscillations that can be easily built and
may be incorporated into an undergraduate physics laboratory, and help students
and teachers to appreciate the complex quantum phenomena of EIT and double EIT
put together in a very simplified manner both theoretically and experimentally.

\section{Model and basic equations}

\begin{figure}
\centering
\includegraphics[scale=0.7]{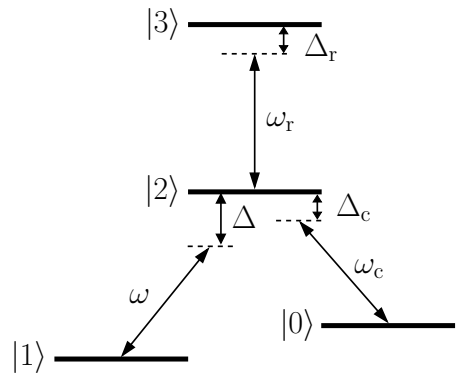}
\caption{\label{Fig01}Schematic energy level diagram of a four-level system in
the inverted-Y configuration. Here $\omega$, $\omega_{\mathrm{c}}$, and
$\omega_{\mathrm{r}}$ are the frequencies of the probe, coupling and pumping
fields, respectively; whereas $\Delta$, $\Delta_{\mathrm{c}}$, and
$\Delta_{\mathrm{r}}$ are their corresponding frequency detunings.}
\end{figure}

We considered a medium consisting of four-level atoms in the so-called
inverted-Y configuration as shown in FIG.~\ref{Fig01}. The levels $\ket{1}$ and
$\ket{2}$ were coupled by a ``probe'' field of frequency $\omega$, in whose
absorption and dispersion we were interested. The level $\ket{2}$ was connected
to the lower level $\ket{0}$ by a strong ``coupling'' field of frequency
$\omega_{\mathrm{c}}$, and to the upper level $\ket{3}$ by the strong
``pumping'' field of frequency $\omega_{\mathrm{r}}$. Only the atomic
transitions $\ket{1}\leftrightarrow\ket{2}$, $\ket{0}\leftrightarrow\ket{2}$,
and $\ket{2}\leftrightarrow\ket{3}$ were dipole allowed.

In a typical double EIT experiment, quantum interference is introduced by
driving the upper two levels with strong coherent fields. Under appropriate
conditions, the medium becomes \textit{transparent} (zero absorption) for the
probe field. In the absence of the coupling and pumping fields, we may observe a
regular absorption resonance profile. However, under certain conditions, the
addition of either the coupling or pumping fields prevents the absorption of
energy by the medium, and the transmitted intensity as a function of the probe
frequency shows a narrow peak of induced transparency called single EIT (or just
EIT). When both coupling and pumping fields are simultaneously present, then
they together control the absorption and propagation of the probe field, and
thus double EIT may be observed in the transmitted intensity profile of the
probe field~\cite{Petrosyan:65}.

\begin{figure}[t]
\centering
\vspace*{0.6cm}
\includegraphics[scale=0.52]{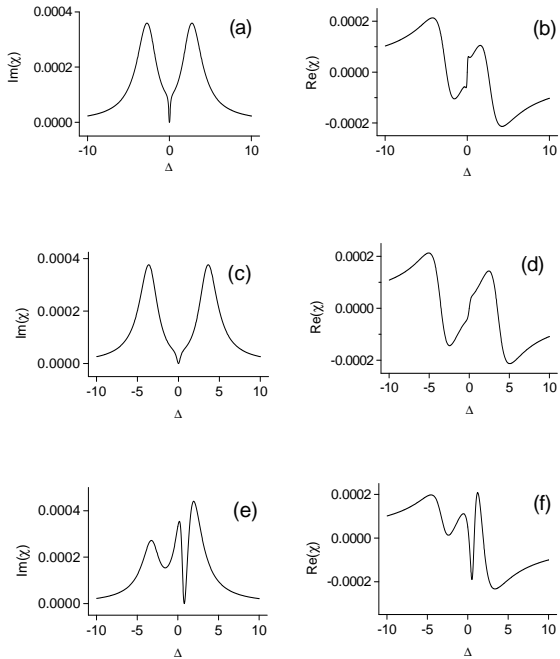}
\caption{\label{Fig02} Imaginary and real parts of the susceptibility $\chi$, as
a function of probe detuning $\Delta$, for different parametric conditions. The
profiles show double EIT for a four-level atom in an inverted-Y configuration
with $\gamma_1=\gamma_2=\gamma_3=1.0$, and $\gamma_0=10^{-4}$. The other
parameters for plots (a,b), (c,d), and (e,f) are ($\Omega_{\mathrm{c}}=1.0$,
$\Omega_{\mathrm{r}}=2.5$, $\Delta_{\mathrm{c}}=0$, $\Delta_{\mathrm{r}}=0$),
($\Omega_{\mathrm{c}}=2.0$, $\Omega_{\mathrm{r}}=3.0$, $\Delta_{\mathrm{c}}=0$,
$\Delta_{\mathrm{r}}=0$), and ($\Omega_{\mathrm{c}}=1.0$,
$\Omega_{\mathrm{r}}=2.0$, $\Delta_{\mathrm{c}}=0.8$,
$\Delta_{\mathrm{r}}=1.8$), respectively. All the parameters have the dimension
of frequency.}
\end{figure}

Absorptive and dispersive properties of the atomic system can be studied by
calculating the electrical susceptibility of the system. When the atom-field
interaction is determined by the density matrix equation and the corresponding
off-diagonal (coherence) component of the probe transition is $\rho_{12}$, the
complex susceptibility $\chi$ is given by $\chi=\mu_{\mathrm{p}}
\rho_{12}/E_{\mathrm{p}}$, in which $\mu_{\mathrm{p}}$ and $E_{\mathrm{p}}$
represent the dipole moment and the field amplitude for the probe transition.
The susceptibility $\chi=\chi' + i \chi''$ is a complex quantity such that its
real (imaginary) part determines the dispersive (absorptive) property of the
atomic medium for the probe field. The intensities of the driving fields
determine the effects observed in double EIT, as depicted in FIG.~\ref{Fig02}
for the radiative decay constants $\gamma_1=\gamma_2=\gamma_3=1.0$, and
$\gamma_0=10^{-4}$. The Rabi frequencies $\Omega_{\mathrm{c}}$ and
$\Omega_{\mathrm{r}}$ are directly proportional to the coupling and pumping
field strengths, respectively, and must be comparable with all damping rates
$\gamma_i$ present in the medium.

Figure~\ref{Fig02}(a) clearly shows double EIT (two dips at $\Delta=0$) at the
exact resonance conditions of the coupling and pumping fields, i.e.,
$\Delta_{\mathrm{c}}=\Delta_\mathrm{r}=0$ and ($\Omega_{\mathrm{c}}=1.0$,
$\Omega_{\mathrm{r}}=2.5$). The corresponding dispersive property is given in
FIG.~\ref{Fig02}(b). Furthermore, strong coupling and pumping fields  may induce
AC-Stark splitting of the excited levels $\ket{2}$ and $\ket{3}$ under resonant
conditions. When the coupling and pumping fields are strong, the splitting
expands, and the absorption spectrum displays the Autler-Townes
doublets\cite{Autler:703}. In FIG.~\ref{Fig02}(c), the values of the coupling
and pumping fields Rabi frequencies are stronger ($\Omega_{\mathrm{c}}=2.0$,
$\Omega_{\mathrm{r}}=3.0$ with other parameters unchanged) in comparison to
FIG.~\ref{Fig02}(a) and hence the width of the two EIT dips becomes broader due
to a wider splitting of the Autler-Townes doublets. The corresponding dispersive
properties under this parametric condition are displayed in FIG.~\ref{Fig02}(d).
Finally, the effect of off-resonant coupling and pumping fields
($\Omega_{\mathrm{c}}=1.0$, $\Omega_{\mathrm{r}}=2.0$,
$\Delta_{\mathrm{c}}=0.8$, $\Delta_{\mathrm{r}}=1.8$) are displayed in
FIG.~\ref{Fig02}(e,f).

Two EIT dips moved away from the $\Delta=0$ position because of finite detunings
of the coupling and pumping fields. The details of this theoretical work on
double EIT are discussed in reference~\cite{Joshi:370}. The single EIT observed
experimentally in a three-level $\Lambda$-type atomic system, and double EIT in
a four-level tripod type atomic system are shown in figure 4 of
Ref.~\onlinecite{Li:43} and figure 3(a1) of Ref.~\onlinecite{Li-S:44},
respectively. Clearly in the optical spectrum, a single deep is observed in the
absorption spectrum of a single EIT system and two deeps are observed in a
double EIT system. The experimental conditions are mentioned in the captions of
the figures and can be further explored in those references.

\subsection{Mechanical spring analog of single and double EIT-like phenomena}

The Lorentz model~\cite{Allen-Eberly,Lorentz} is recognized as one of the
classical models for the atom that works incredibly well for describing the
interaction of light with matter. The basic assumption made in this model is
that the bounded electrons within the neutral atom oscillate about their
equilibrium position with a very small amplitude. In addition, each electron-ion
pair behaves as a simple harmonic oscillator which couples to the
electromagnetic field through its electric dipole moment. Thus, the atom can be
described as a damped harmonic oscillator of mass $m$ attached to a rigid
support by a spring of force constant $\kappa$ and driven by a harmonic force
$F=F_0 e^{- i(\omega t+\phi)}$. The inclusion of a damping force is necessary in
the model because different physical processes, like atomic collisions and
radiative decays, may take away energy from the atom. The forces acting on the
oscillator (with its natural frequency $\omega_0=\sqrt{\kappa/m}$) are the
harmonic driving force $F$, the spring force $-\kappa x$, and the damping force
$2\beta\dot{x}$, where $\dot{x}$ is the oscillator's speed. Newton's second law
gives the equation of motion for the position variable $x$ in the Lorentz model
as
\begin{equation}\label{Eq:Lorentz}
\ddot{x}+2\beta\dot{x}+\omega_0^{2}x= F_0 e^{- i(\omega t+\phi)}\,.
\end{equation}

\begin{figure}
\centering
\includegraphics[scale=0.6]{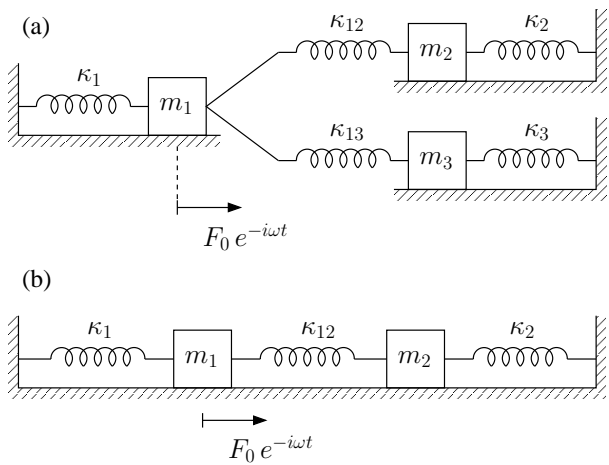}
\caption{\label{Fig05}Coupled damped harmonic-oscillator model
showing (a) double EIT and (b) single EIT features.}
\end{figure}

In the classical model of double EIT, we described the atom as a damped harmonic
oscillator of mass $m_1$ attached to a rigid support by a spring of force
constant $\kappa_1$ and driven by a harmonic force $F = F_0\, e^{- i(\omega t +
\phi)}$. To this mass-spring combination were attached two other masses
originally at rest, $m_2$ and $m_3$ that were connected to mass $m_1$ by springs
of force constants $\kappa_{12}$ and $\kappa_{13}$, respectively. These two
masses were also fixed, from the other side, to rigid supports by springs of
force constants $\kappa_2$ and $\kappa_3$, respectively [see
FIG.~\ref{Fig05}(a)].

It is always a matter of importance and interest to know at what rate energy is
transmitted into the driven oscillator, and how this power is absorbed as a
function of the frequency $\omega$~\cite{French}. In the typical situation of a
damped harmonic oscillator $m_1$ driven by a harmonic force $F$, a standard
absorption resonance profile is observed. However, if either $m_2$ or $m_3$ is
allowed to move due only to the forces from the springs they are attached to
(with force constants $\kappa_{12}$ and $\kappa_2$, and $\kappa_{13}$ and
$\kappa_3$, respectively), this will avoid absorption in a limited region of the
resonance profile, and the transmitted power as a function of the driving force
frequency will show a narrow peak of induced transparency (single
EIT)\cite{Garrido:37}.

In this physical model of the atom, the spring attaching masses $m_1$ and $m_2$
(with force constant $\kappa_{12}$) emulated the coupling field between atomic
levels $\ket{0}$ and $\ket{2}$, whereas the spring connecting masses $m_1$ and
$m_3$ (with force constant $\kappa_{13}$) emulated the pumping field between
levels $\ket{2}$ and $\ket{3}$. The probe field was then modeled by the harmonic
force acting on mass $m_1$. These analogues remind us the description of the
fields in terms of harmonic oscillators~\cite{Scully-Zubairy}. Now, if we allow
\textit{both} masses $m_2$ and $m_3$ to move simultaneously under the conditions
described above, we will observe double EIT features.

To describe the classical evolution of this system, we used a fixed set of
one-dimensional Cartesian coordinates $x_1$, $x_2$, and $x_3$, representing the
positions of the masses from their equilibrium positions. Thus, the equations of
motion could be written like
\begin{align}\label{Eq:xi}
\begin{split}
& \ddot{x}_1(t) + \gamma_1\,\dot{x}_1(t) + \omega_1^2\,x_1(t)
\\
& \hspace{1.45em}- \Omega_{\mathrm{c}}^2\,x_2(t) - \Omega_{\mathrm{r}}^2\,x_3(t)
= (F_0/m)\, e^{- i\omega t},
\\
& \ddot{x}_2(t) + \gamma_2\,\dot{x}_2(t) + \omega_2^2\,x_2(t)
- \Omega_{\mathrm{c}}^2\,x_1(t) = 0,
\\
& \ddot{x}_3(t) + \gamma_3\,\dot{x}_3(t) + \omega_3^2\,x_3(t)
- \Omega_{\mathrm{r}}^2\,x_1(t) = 0,
\end{split}
\end{align}
where we assumed that $\phi=0$ and $m_1 = m_2 = m_3 \equiv m$. The other
parameters were defined as follows:
$\omega_1^2=(\kappa_1+\kappa_{12}+\kappa_{13})/m$,
$\omega_2^2= (\kappa_2+\kappa_{12})/m$, $\omega_3^2 =(\kappa_3+\kappa_{13})/m$,
$\Omega_{\mathrm{c}}^2= \kappa_{12}/m$, and
$\Omega_{\mathrm{r}}^2=\kappa_{13}/m$. The damping parameters $\gamma_i$
(viscous damping) represented the mechanical equivalent to the spontaneous decay
rates of the three excited states in the inverted-Y atomic configuration.

Because we expected the motion to be oscillatory, we attempted solutions of the
form $x_i = B_i\, e^{- i \omega t}$, with $B_i$s are constants ($i=1,\,2,\,3$).
Substituting these expressions for the displacements into the equations of
motion, we found that the displacement of $m_1$ (atom displacement) was given by
\begin{widetext}
\begin {equation}\label{Eq:x1}
x_1(t) = \dfrac{(F_0/m)\, e^{- i \omega t}} {(\omega_1^2 - \omega^2 -
i\,\gamma_1\,\omega) - \dfrac{\Omega_{\mathrm{c}}^4}{\omega_2^2 - \omega^2
-i\,\gamma_2\,\omega} - \dfrac{\Omega_{\mathrm{r}}^4}{\omega_3^2 - \omega^2 -
i\,\gamma_3\,\omega}}\,.
\end{equation}
\end{widetext}

In the Lorentz oscillator model\cite{Lorentz,Allen-Eberly}, the electrical
polarization $\mathcal{P}$ (or the susceptibility $\chi=\mathcal{P}/F$) induced
in the atom by the external force field $F$ is directly proportional to $x_1$,
for the polarization is defined as $\mathcal{P} = N\,e\,x_1$, where $N$ is the
number of atoms per unit volume, and $e$ is the electronic charge. The real and
imaginary parts of $x_1$ give the dispersion  and absorption properties of the
atom, respectively. A graphical analysis of~\eqref{Eq:x1} will allow us to
explore these two important properties of light propagation. The frequency
differences (detuning) of the probe, coupling, and pumping fields with respect
to the external driving field were defined like $\Delta = \omega_1 - \omega$,
$\Delta_{\mathrm{c}} = \omega_2 - \omega$, and $\Delta_{\mathrm{r}} = \omega_3 -
\omega$, respectively. These definitions are slightly different from what is
used in optical double EIT nomenclature~\cite{Joshi:370}.

\subsection{Electrical analog of double EIT: coupled RLC circuits}

\begin{figure}
\centering
\includegraphics[scale=0.7]{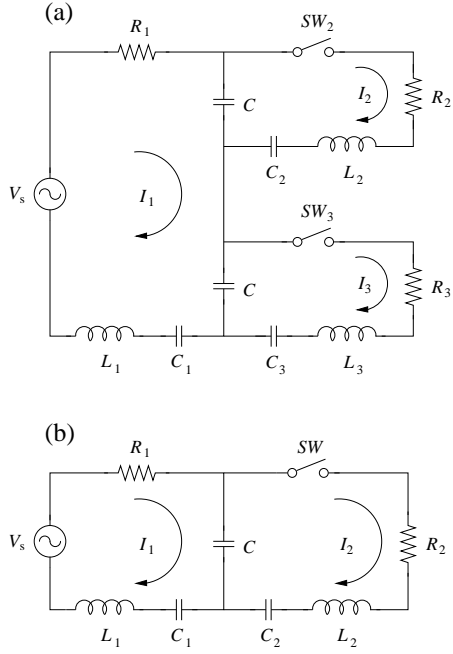}
\caption{\label{Fig06}Coupled RLC electrical circuits displaying (a)
double EIT-like and (b) single EIT-like features.}
\end{figure}

There is a well known correspondence between a driven damped harmonic oscillator
and an electrical circuit consisting of a resistor $R$, an inductor $L$, and a
capacitor $C$ connected in series to an alternating voltage source
$V$\cite{Marion-Thornton}. The importance of this correspondence is that RLC
circuits are easy to build in the laboratory, and may be used as excellent
examples of \textit{non mechanical} oscillations. We used these circuits to
demonstrate experimentally and study theoretically single and double EIT by
analyzing the dissipation of electric power in the resistance. The circuit that
showed double EIT behavior is shown in FIG.~\ref{Fig06}(a). This circuit was
made up of three loops of RLC circuits. The resistance, inductance, and
capacitance of the loops were represented by $R_i$, $L_i$, and $C_i$,
respectively ($i=1,\,2,\,3$). The first loop with resistance $R_1$, inductance
$L_1$, capacitances $C_1$ and $C/2$, represented the atom. The resistance
accounted for the spontaneous radiative decay of the second excited level
$\ket{2}$ to level $\ket{1}$. The capacitance $C$, shared by the first and
second loops, provided the link between the atom and the coupling field; whereas
the other capacitance $C$, shared by the first and third loops, linked the atom
with the pumping field.

In this circuit, the loop that modeled the atom (loop 1) had a resonance
frequency that represented the transition energy from the ground state $\ket{1}$
to the excited state $\ket{2}$. The probability of populating this excited state
was a maximum when the alternating voltage source $V$ was in resonance with the
resonance frequency of this loop, (or in resonance with the $\ket{1} \rightarrow
\ket{2}$ transition). However, with a three-loop configuration, we had two other
possible ways to accomplish this excitation since we were using the analogue of
a four-level atom in the inverted-Y configuration. For instance, loop 1
(representing the atom) could also have been excited either by the
\textit{coupling} loop 2 ($\ket{0} \rightarrow \ket{2}$), the \textit{pumping}
loop 3 ($\ket{3}\rightarrow \ket{2}$), or both.

The EIT was studied by examining the frequency dependence of the transmitted
power from the voltage source $V=V_{\mathrm{s}}\, e^{- i \omega t}$ to the
resonant first loop. If the currents flowing in the three different loops of the
circuit are written like $I_1(t)=\dot{q}_1(t)$, $I_2(t)=\dot{q}_2(t)$, and
$I_3(t)=\dot{q}_3(t)$, the following system of coupled differential equations
for the charges is found
\begin{align}\label{Eq:currents}
\begin{split}
& \ddot{q}_1(t) + \gamma_1\,\dot{q}_1(t) + \omega_1^2\,q_1(t)
\\
& \hspace{1.05em}-\Omega_{\mathrm{c}}^2\,q_2(t) - \Omega_{\mathrm{r}}^2\,q_3(t)
= (V_{\mathrm{s}}/L_1)\, e^{- i \omega t},
\\
& \ddot{q}_2(t) + \gamma_2\,\dot{q}_2(t) + \omega_2^2\,q_2(t) -
\Omega_{\mathrm{c}}^2\,q_1(t) = 0,
\\
& \ddot{q}_3(t) + \gamma_3\,\dot{q}_3(t) + \omega_3^2\,q_3(t) -
\Omega_{\mathrm{r}}^2\,q_1(t) = 0,
\end{split}
\end{align}
where $\gamma_i= R_i/L_i$, $\omega_i^2=1/(L_i C_{\mathrm{e}i})$ (with
$i=1,\,2,\,3$), and $\Omega_{\mathrm{c}}^2 = \Omega_{\mathrm{r}}^2 = 1/(L_1C)$.
The equivalent capacitances for these loops were
\begin{align}\label{Eq:Ce3}
\begin{split}
C_{\mathrm{e}1} &= \dfrac{(C/2)\,C_1}{C/2 + C_1},
\\
C_{\mathrm{e}2} &= \dfrac{C\,C_2}{C + C_2},
\\
C_{\mathrm{e}3} &= \dfrac{C\,C_3}{C + C_3}.
\end{split}
\end{align}
It was easy to compare~\eqref{Eq:xi} with~\eqref{Eq:currents} and conclude that
both models described the same physical phenomenon.

Applying the Kirchhoff's second law to the three loops of the
circuit\cite{Symon}, with loop currents $I_1$, $I_2$, and $I_3$, we obtained
\begin{align}\label{Eq:Kirchhoff}
\begin{split}
& [R_1 -  i(2 X_C + X_{C_1} - X_{L_1})] I_1
\\
& \hspace{8.5em} +  i X_C I_2 +  i X_C I_3 = V,
\\
&  i X_C I_1 + [R_2 -  i(X_C + X_{C_2} - X_{L_2})] I_2 = 0,
\\
&  i X_C I_1 + [R_3 -  i(X_C + X_{C_3} - X_{L_3})] I_3 = 0,
\end{split}
\end{align}
where $X_{C}=1/(\omega\,C)$ and $X_{C_i}=1/(\omega\,C_i)$ ($i=1,\,2,\,3$) were
the capacitive reactances, and $X_{L_i}=\omega L_i$ ($i=1,\,2,\,3$) were the
inductive reactances. From the above system of equations, it was found that
\begin{equation}\label{Eq:I2}
I_1 = \left(\dfrac{A +  i B}{A^2 + B^2}\right) V,
\end{equation}
where, for convenience, we defined
\begin{widetext}
\begin{align}
&A \equiv R_1 + \dfrac{R_2 X_C^2}{R_2^2 + [X_{L_2} - (X_C+X_{C_2})]^2} +
\dfrac{R_3 X_C^2}{R_3^2 + [X_{L_3}-(X_C + X_{C_3})]^2},\label{Eq:AB1}
\\[0.8em]
&B \equiv X_{L_1} - (2X_C + X_{C_1}) - \dfrac{X_C^2 [X_{L_2}-(X_C +
X_{C_2})]}{R_2^2 + [X_{L_2}-(X_C + X_{C_2})]^2}
- \dfrac{X_C^2 [X_{L_3}-(X_C + X_{C_3})]}{R_3^2 + [X_{L_3}-(X_C + X_{C_3})]^2}.
\label{Eq:AB2}
\end{align}
\end{widetext}
The electrical power in the $R_1 L_1 C_{\mathrm{e}1}$ loop was obtained by
multiplying~\eqref{Eq:I2} by the voltage. In-phase and out-of-phase components
of the power were associated with the energy dissipated by the resistive
($P_{\mathrm{R}}$), and the energy stored by the reactance ($P_{\mathrm{X}}$)
parts of the circuit, giving the following expressions
\begin{equation}
P_{\mathrm{R}} = \dfrac{A\,|V_{\mathrm{s}}|^2}{A^2 + B^2}\, \qquad \mbox{and}
\qquad
P_{\mathrm{X}} = \dfrac{B\,|V_{\mathrm{s}}|^2}{A^2 + B^2},
\end{equation}
where $A$ and $B$ were given by~\eqref{Eq:AB1} and~\eqref{Eq:AB2}.

\begin{figure}[t]
\centering
\includegraphics[scale=1.0]{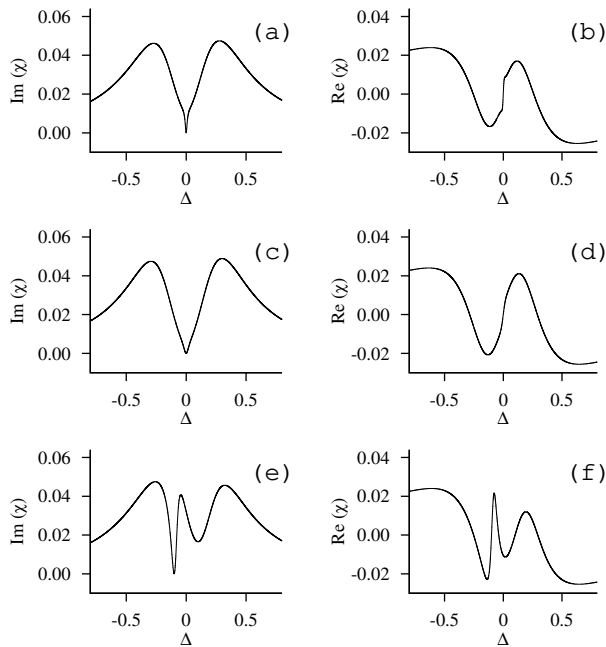}
\caption{\label{Fig07} Imaginary and real parts of the susceptibility $\chi$ for
the double EIT-like system as a function of the detuning $\Delta $. The
radiative decays are $\gamma_1=1.0$, $\gamma_2=0.1$, $\gamma_3=10^{-4}$, and the
coupling and pumping detunings $\Delta_{\mathrm{c}} = \Delta_{\mathrm{r}}=0$.
For plots (a) and (b): $\Omega_{\mathrm{c}}=3.0$, and $\Omega_{\mathrm{r}}=2.3$.
For plots (c) and (d): $\Omega_{\mathrm{c}}=2.7$ and $\Omega_{\mathrm{r}}=3.0$.
Plots (e) and (f) have $\Delta_{\mathrm{c}}=\Delta_{\mathrm{r}}=0.1$ and
$\Omega_{\mathrm{c}}=3.0$ and $\Omega_{\mathrm{r}}=2.3$. All the parameters have
the dimension of frequency.}
\end{figure}

We first studied the absorption and dispersion properties of the spring-mass
system at exact resonance conditions $\Delta_{\mathrm{c}}=\Delta_{\mathrm{r}}=0$
of the coupling and pumping fields. Figures~\ref{Fig07}(a) and~\ref{Fig07}(b)
display the curves for the absorption and dispersion of the probe field,
respectively. The Rabi frequencies and radiative decays (damping) used were
$\Omega_{\mathrm{c}} = 3.0$, $\Omega_{\mathrm{r}} = 2.3$, $\gamma_1 = 1.0$,
$\gamma_2 = 0.1$, and $\gamma_3 = 10^{-4}$ (all these quantities given in units
of the atomic decay $\gamma_1$). Double EIT was observed in the absorption curve
at $\Delta = 0$, where two dips of different widths, one inside the other,
clearly became visible [FIG.~\ref{Fig07}(a)]. When the coupling and pumping
frequencies were changed to $\Omega_{\mathrm{c}} = 2.7$ and $\Omega_{\mathrm{r}}
= 3.0$, we noticed from the absorption curve that when the pumping field
increased reducing its relative difference with the coupling field, the second
dip became wider [FIG.~\ref{Fig07}(c)]. On the other hand, FIG.~\ref{Fig07}(d)
shows how, in the vicinity of $\Delta = 0$, the peaks of dispersion flipped in a
smoother way. The change of frequency detunings brought in further interesting
changes as depicted in FIG.~\ref{Fig07}(e) and~\ref{Fig07}(f), where we set
$\Delta_{\mathrm{c}} = \Delta_{\mathrm{r}} = 0.1$. Because of these detuning
changes, the two EIT peaks separated from each other, and moved away relative to
the $\Delta = 0$ position.

\begin{figure}[t]
\centering
\includegraphics[scale=1.0]{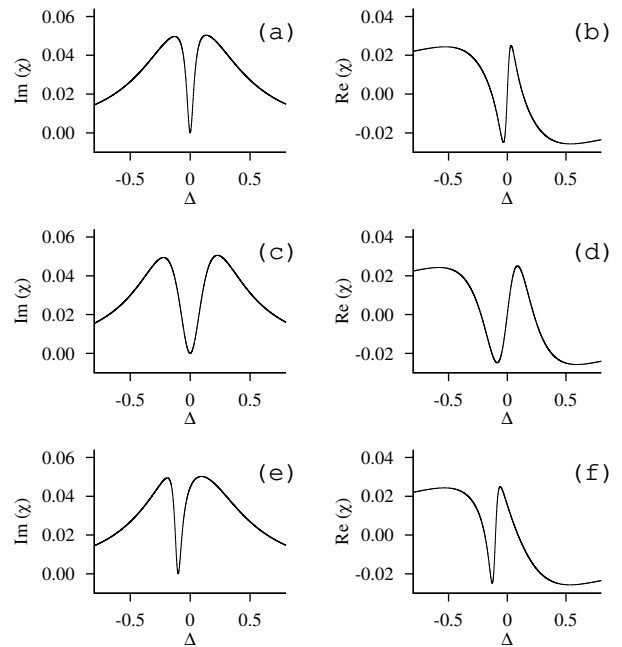}
\caption{\label{Fig08} Imaginary and real parts of the susceptibility $\chi$ for
the single EIT-like system as a function of the detuning $\Delta $. The
radiative decays are $\gamma_1=1.0$, $\gamma_2=10^{-4}$, and $\gamma_3=0.0$. For
plots (a) and (b), $\Delta_{\mathrm{c}}=0.0$ and $\Omega _{\mathrm{c}}=2.3$. For
plots (c) and (d), $\Delta_{\mathrm{c}}=0.0$ and $\Omega_{\mathrm{c}}=3.0$.
Plots (e) and (f), $\Delta_{\mathrm{c}}=0.1$ and $\Omega_{\mathrm{c}}=2.3$. All
the parameters have the dimension of frequency.}
\end{figure}

The double EIT features changed to those of single EIT [for the spring-mass
system in FIG.~\ref{Fig05}(b)] when the zero-limit condition for either the
coupling or pumping fields were considered (i.e., $\Omega_{\mathrm{c}}=0$ or
$\Omega_{\mathrm{r}}=0$). The absorption and dispersion curves showed
characteristics of standard EIT, as observed in FIG.~\ref{Fig08}(a)
and~\ref{Fig08}(b), respectively. The parameters used for this case were
$\Omega_{\mathrm{c}}=2.3$, $\Omega_\mathrm{r}=0.0$, $\gamma_1=1.0$,
$\gamma_2=10^{-4}$, and $\gamma_3=0.0$.
The effects of the coupling field strength on this system are now shown in
FIG.~\ref{Fig08}(c) and~\ref{Fig08}(d). The only parameter changed was
$\Omega_{\mathrm{c}}=3.0$. The broadening in the EIT peak was apparent and
caused by the coupling field increase. In FIG.~\ref{Fig08}(e)
and~\ref{Fig08}(f), the only parameter changed was $\Delta _{\mathrm{c}}= 0.1$,
leaving the other parameters as before. Clearly, the EIT moved away from the
centre of the graph.

\begin{figure}[t]
\centering
\includegraphics[scale=0.75]{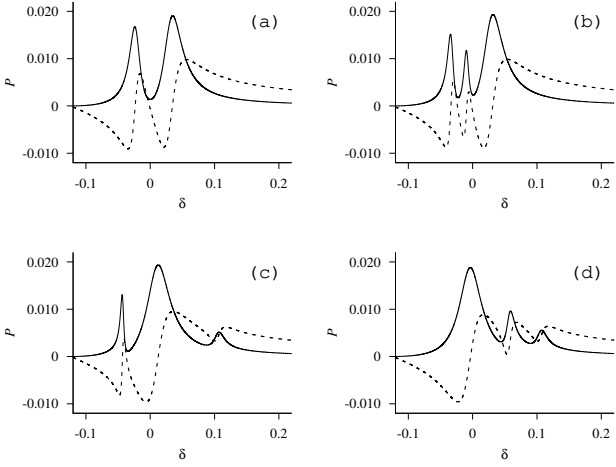}
\caption{\label{Fig09}Power transferred to the $R_1 L_1 C_{\mathrm{e}1}$
circuit (Fig.6(a)) as a function of the detuning $\delta \equiv \omega -
\omega_{\mathrm{R}}$. This detuning was defined as the difference between the
driving field frequency $\omega$ and the resonance frequency of the circuit
$\omega_{\mathrm{R}}$. The parameters used were $R_2 = R_3 =
5.0\,\Omega$, $R_1 = 50\,\Omega$, $C_1 = C_2 = C_3 =
0.1\,\mu\mbox{F}$, $C = 0.2\,\mu\mbox{F}$, and $L_1 =
0.0010\,\mbox{H}$. For plots (a) $L_2 = 0.0010\,\mbox{H}$ and $L_3 =
0.0010\,\mbox{H}$; (b) $L_2 = 0.0010\,\mbox{H}$ and $L_3 =
0.0015\,\mbox{H}$; (c) $L_2 = 0.0020\,\mbox{H}$ and $L_3 =
0.0003\,\mbox{H}$; and (d) $L_2 = 0.0005\,\mbox{H}$ and $L_3 =
0.0003\,\mbox{H}$. The solid line represents $P_{\mathrm{R}}$, whereas the
dashed line represents $P_{\mathrm{X}}$. $P$ is given in arbitrary units.}
\end{figure}

We next looked at the behavior of $P_\mathrm{R}$ and $P_\mathrm{X}$ as a
function of the frequency detuning $\delta = \omega - \omega_{\mathrm{R}}$ for
different initial conditions of the parameters $R$, $L$, and $C$. In
FIG.~\ref{Fig09}, the effects of the coupling and pumping frequency detunings,
in the double EIT scenario, are shown when parameters $L_2$ and $L_3$ took on
different values. The solid and dashed lines represent the absorption and
dispersion of light, respectively. Figure~\ref{Fig09}(a) shows that, at exact
resonance conditions for both the coupling and pumping fields with the probe
field ($\Delta_{\mathrm{c}}=\Delta_{\mathrm{r}}=0$), there was only a single dip
in the curve (like single EIT). This happened because both EIT dips occurred at
the same location. The corresponding dispersion curve also shows this particular
characteristic. A separation of the two EIT dips in the absorption line occurred
when $L_3$ was increased, as shown in FIG.~\ref{Fig09}(b). The dispersion line
also moved apart, showing the typical dispersion characteristics of double EIT.
This showed how the second dip moved toward the left in comparison to the one
displayed in plot~\ref{Fig09}(a). The separation of the two EIT dips is shown
more clearly in FIG.~\ref{Fig09}(c) for a different set of parameters $L_2$ and
$L_3$. The two dips moved in opposite directions, and double EIT was visible
again. The dispersion curve also showed double EIT, and the peaks moved in
opposite directions. Figure~\ref{Fig09}(d) shows the dips shifted to the right
for yet another different set of parameters $L_2$ and $L_3$.

\begin{figure}
\centering
\includegraphics[scale=0.75]{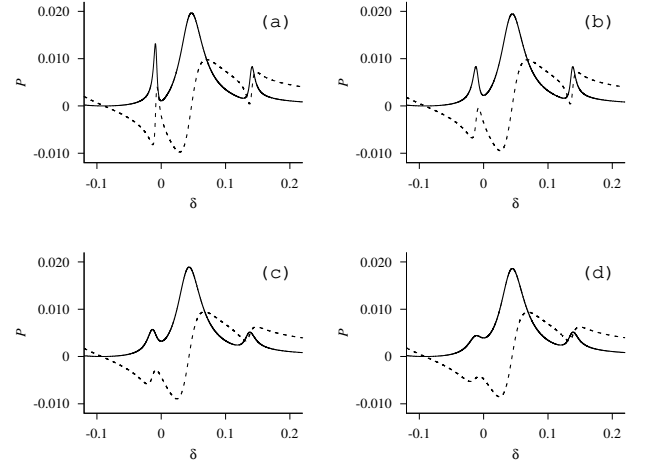}
\caption{\label{Fig10} Power transferred to the $R_1 L_1 C_{\mathrm{e}1}$
circuit (Fig.4(a)) as a function of the detuning $\delta$. The parameters used
were $R_1 = 50\,\Omega$, $L_2 = 0.0020\,\mbox{H}$, $L_1 = 0.0010\,\mbox{H}$,
$L_3 = 0.0003\,\mbox{H}$, $C_1 = C_2 = C_3 =0.1\,\mu\mbox{F}$, and $C =
0.2\,\mu\mbox{F}$. For plots (a) $R_2 = 5.0\,\Omega$ and $R_3 = 2.0\,\Omega$;
(b) $R_2 = 15\,\Omega$ and $R_3 = 2.0\,\Omega$; (c) $R_2 = 30\,\Omega$ and $R_3
= 5.0\,\Omega$; and (d) $R_2 = 50\,\Omega$ and $R_3 = 5.0\,\Omega$.
respectively. The solid line represents $P_\mathrm{R}$, whereas the dashed line
represents $P_\mathrm{X}$. $P$ is given in arbitrary units.}
\end{figure}

Different values of the radiative decay parameters (damping) also changed the
absorption and dispersion curves in double EIT. In the electrical analogue of
the atom, the resistance in the circuit loops represented the damping. A
comparison of FIG.~\ref{Fig10}(a) and~\ref{Fig10}(b) shows how the first EIT dip
became less pronounced, and its width expanded when resistance $R_2$ increased.
When we increased $R_2$ and $R_3$ even more, both EIT dips became even less
pronounced, and their widths increased with the resistance increase [compare
FIG.~\ref{Fig10}(a) and~\ref{Fig10}(c)]. A large value of $R_2$ caused the first
dip to spread out increasing its width and decreasing its depth.

\begin{figure}
\centering
\includegraphics[scale=0.75]{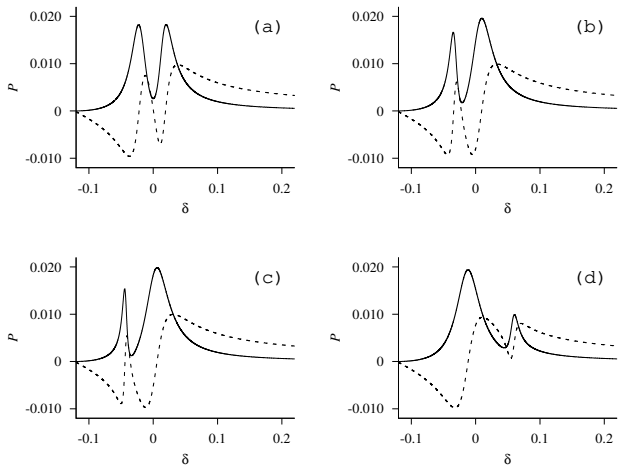}
\caption{\label{Fig11} Power transferred to the $R_1 L_1 C_{\mathrm{e}1}$
circuit (Fig.4(b)) as a function of the detuning $\delta$. The parameters used
were $R_2 = 5.0\,\Omega$, $R_1 = 50\,\Omega$, $C_1 = C_2 = 0.1\,\mu\mbox{F}$, $C
= 0.2\,\mu\mbox{F}$, and $L_1 = 0.0010\,\mbox{H}$. For plots (a) $L_2 =
0.0010\,\mbox{H}$, (b) $L_2 = 0.0015\,\mbox{H}$, (c) $L_2 = 0.0020\,\mbox{H}$,
and (d) $L_2 = 0.0005\,\mbox{H}$. The solid line represents $P_R$, whereas the
dashed line represents $P_X$. $P$ is given in arbitrary units.}
\end{figure}

By removing one of the loops, i.e., either $R_2 L_2 C_{\mathrm{e}2}$ or $R_3 L_3
C_{\mathrm{e}3}$, we recovered the two RLC coupled circuits showing single EIT
(see FIG.~\ref{Fig06}). Figure~\ref{Fig11}[(a)--(d)] show the behavior of
$P_\mathrm{R}$ and $P_\mathrm{X}$ after disconnecting the pumping loop $R_3 L_3
C_{\mathrm{e}3}$. Plot~\ref{Fig11}(a) clearly shows a single EIT dip at exact
resonance. The frequency detunings of the two RLC loops were zero; the two loops
had the same resonance frequency determined from the selected parameters of each
loop. By changing the value of $L_2$, the resonance frequencies of the two loops
changed, and so did the absorption and dispersion curves, as shown in
FIG.~\ref{Fig11}(b). The symmetry of the curves was lost because of the
frequency detunings of the two circuits. A further increase in the $L_2$ value
shifted the EIT dip even further [compare FIG.~\ref{Fig11}(c)
with~\ref{Fig11}(b)]. In contrast, when the value of $L_2$ was decreased
relative to $L_1$, the EIT dip moved in the opposite direction [
FIG.~\ref{Fig11}(d)] as if there were a negative frequency detuning between the
two circuit loops in comparison to plots~\ref{Fig11}(b) and~\ref{Fig11}(c).

On the other hand, the experimental results obtained for the coupled RLC circuit
shown in FIG.~\ref{Fig06}(b) displayed single EIT behavior. We measured the
current flowing through the resistor $R_1$ and calculated the power delivered to
the $R_1 L_1 C_{\mathrm{e}1}$ loop. Figure~\ref{Fig10} shows the power
transmitted $P_\mathrm{R}$ as a function of the driving field frequency
$\omega$.

\begin{figure}[t]
\centering
\vspace*{0.6cm}
\includegraphics[scale=0.43]{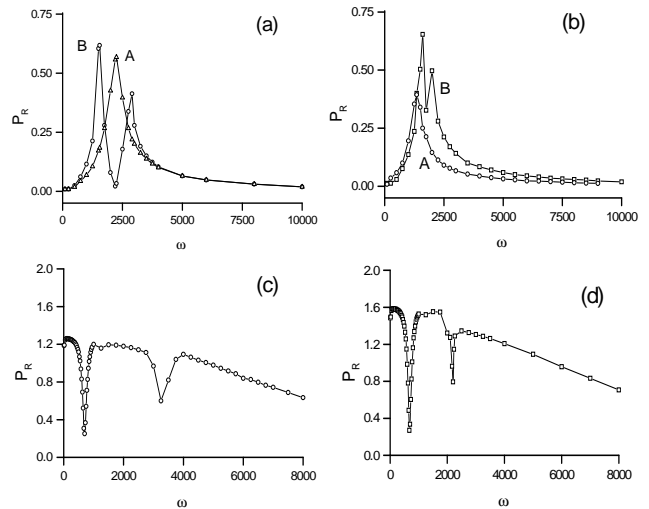}
\caption{\label{Fig12} Experimental plots of the power transferred $P_R$ to the
$R_1 L_1C_{\mathrm{e}1}$ loop as a function of the driving field frequency
$\omega$. Plots (a) and (b) show single EIT-like behavior, whereas (c) and (d)
show double EIT. The curves A and B depict the situation when switch \textit{SW}
is open and closed, respectively.  The parameters for this circuit are
$R_1=50\,\Omega$, $R_2 = 5.0\,\Omega$, $C_1 = C_2 =0.10\,\mu\mbox{F}$, and $L_1
= L_2 = 0.27\,\mbox{H}$. For plots (a) $C = 0.047\,\mu\mbox{F}$, (b) $C =
0.2\,\mu\mbox{F}$. For plots (c) and (d), the parameters of the circuit are $R_1
= 87\,\Omega$, $R_2 = 16.0\,\Omega$, $R_3 = 25\,\Omega$, $C_1 = C_2 = C_3 =
0.047\,\mu\mbox{F}$, $C = 0.1\,\mu\mbox{F}$, and $L_1 = 0.27\,\mbox{H}$. In
particular, for (c) $L_2 = 0.065\,\mbox{H}$, and (d) $L_2= 0.185\,\mbox{H}$,
with $L_1=L_3= 0.27\,\mbox{H}$}
\end{figure}

In FIG.~\ref{Fig12}(a) and~\ref{Fig12}(b), the curves A and B illustrate the
situation when we opened the switch \textit{SW} (driven single RLC circuit), and
when we closed it (driven RLC circuit coupled to a second RLC circuit). With the
open switch, no power was transferred from the circuit loop $R_2 L_2
C_{\mathrm{e}2}$, and the circuit loop $R_1 L_1 C_{\mathrm{e}1}$ behaved like a
simple, driven RLC circuit as shown in the figures. However, with the closed
switch, we clearly observed a dip (curves B of these two plots). This dip
resembled the single EIT-like dip shown in FIG.~\ref{Fig11}[(a)--(d)]. The two B
curves observed in FIG.~\ref{Fig12} (a) and (b), essentially represented
different resonance frequencies for the observed single EIT in the RLC circuits.

Note that such a simulation of single EIT along with experimental demonstrations
have also been presented in an earlier work by Garrido Alzar et
al~\cite{Garrido:37}. They present a classical analogue of EIT using two
coupled harmonic oscillators subject to a harmonic driving force [similar to
FIG.~\ref{Fig05}(b)] and reproduce the phenomenology observed in EIT by changing
the strength of the coherent coupling field. Moreover, these authors also
recreate EIT behavior experimentally using two linearly coupled RLC circuits
[similar to FIG.~\ref{Fig06}(b)]. In their work, the simulations are for the
degenerate probe and coupling transitions, showing excellent agreement of the
theoretical modeling of EIT (using coupled RLC circuits) with experimental
results under similar conditions of parameters. In this paper, we have presented
not only an extensive simulation of EIT, including the effects of different
strengths of coupling fields and frequency detunings associated with the coupled
harmonic oscillators and RLC circuits, but also simulations for a richer
phenomenon of double EIT. We have selected different sets of parameters in the
experimental simulations to show that single EIT [see FIG.~\ref{Fig12}(a,b)] and
double EIT [see FIG.~\ref{Fig12}(c,d) to be discussed in the following
paragraph] are exhibited for a wider range of parameters, and hence they are
quite versatile.

We observed double EIT in the three coupled RLC circuits as shown in
FIG.~\ref{Fig06}(a). When we experimentally measured the power transferred to
the $R_1 L_1 C_{\mathrm{e}1}$ loop from the loops $R_2 L_2 C_{\mathrm{e}2}$ and
$R_3 L_3 C_{\mathrm{e}3}$, two dips were visible [see FIG.~\ref{Fig12} (c) and
(d)]. We also noted that the position of the second EIT peak changed for
different values of the inductance $L_2$. These dips were the analogues of
quantum interference observed in double EIT atomic systems. In this case, the
interference happened because of the power delivered to the resonant $R_1 L_1
C_{\mathrm{e}1}$ circuit from the voltage source $V$ and the other two coupled
circuits $R_2 L_2 C_{\mathrm{e}2}$ and $R_3 L_3 C_{\mathrm{e}3}$. Classically,
we looked at this phenomena as the interference between three excitation paths
corresponding to the normal modes of oscillation of the coupled harmonic
oscillators.

\section{Summary}

We have presented mechanical and electrical analogies for single, and double EIT
observed in three- and four-level atomic systems using coupled harmonic
oscillator models and RLC circuits. The mechanical analogy, consisting of a
coupled spring-mass system, may be helpful in understanding the observed zero
power absorption in single and double EIT phenomena, as a result of destructive
interference between the normal modes of oscillation of the system. The
dissipation rates of the coupling and pumping oscillators ($\gamma_2$ and
$\gamma_3$, respectively) should be small compared with that of the atomic
oscillator ($\gamma_1$) for EIT to be observable. The symmetry of the equation
of motion of the atom for EIT allows us to study easily, the absorption and
dispersion of a multilevel system in the inverted-Y (four-level) and $\Lambda$
(three-level) configurations.

The electrical analogy, associated with a coupled RLC circuit, may be helpful to
realize the single and double EIT phenomena experimentally. This type of circuit
corresponds to the electrical analogue of the mass-spring system. This fact
allows us to establish a direct correspondence between an atomic system (based
on the Lorentz's approximations) and the RLC circuit. In fact, by changing some
circuit parameters like the inductances and capacitances, it is possible to
produce different control fields acting on different atomic transitions. The
resistances of the circuit represent the radiative decays of these atomic
levels. The Rabi frequencies of these control fields should be large enough from
the radiative decays for EIT to be observable.

The interest on these experiments, and the final purpose of this work is to help
undergraduate students to develop a better understanding of single and double
EIT, as well as to improve their experimental skills. These experiments are easy
to adopt in any undergraduate physics laboratory, and can be used to approach
other compelling topics such as quantum coherence and quantum interference,
which occur in atomic systems, and are particularly important in observing
phenomena like group velocity reduction of light, superconductivity and
superfluidity, and quantum information processing.

\begin{acknowledgments}
The authors gratefully acknowledge the Research Corporation, the College of
Sciences and the Department of Physics at Eastern Illinois University, and the
School of Mathematical and Natural Sciences at the University of
Arkansas-Monticello for providing funding and support for this work.
\end{acknowledgments}

\bibliography{doubleEIT.bib}

\begin{thebibliography}{10}%
\makeatletter
\providecommand \@ifxundefined [1]{%
 \ifx #1\undefined \expandafter \@firstoftwo
 \else \expandafter \@secondoftwo
\fi
}%
\providecommand \@ifnum [1]{%
 \ifnum #1\expandafter \@firstoftwo
 \else \expandafter \@secondoftwo
\fi
}%
\providecommand \enquote [1]{``#1''}%
\providecommand \bibnamefont  [1]{#1}%
\providecommand \bibfnamefont [1]{#1}%
\providecommand \citenamefont [1]{#1}%
\providecommand\href[0]{\@sanitize\@href}%
\providecommand\@href[1]{\endgroup\@@startlink{#1}\endgroup\@@href}%
\providecommand\@@href[1]{#1\@@endlink}%
\providecommand \@sanitize [0]{\begingroup\catcode`\&12\catcode`\#12\relax}%
\@ifxundefined \pdfoutput {\@firstoftwo}{%
 \@ifnum{\z@=\pdfoutput}{\@firstoftwo}{\@secondoftwo}%
}{%
 \providecommand\@@startlink[1]{\leavevmode\special{html:<a href="#1">}}%
 \providecommand\@@endlink[0]{\special{html:</a>}}%
}{%
 \providecommand\@@startlink[1]{%
  \leavevmode
  \pdfstartlink
   attr{/Border[0 0 1 ]/H/I/C[0 1 1]}%
   user{/Subtype/Link/A<</Type/Action/S/URI/URI(#1)>>}%
  \relax
 }%
 \providecommand\@@endlink[0]{\pdfendlink}%
}%
\providecommand \url  [0]{\begingroup\@sanitize \@url }%
\providecommand \@url [1]{\endgroup\@href {#1}{\urlprefix}}%
\providecommand \urlprefix [0]{URL }%
\providecommand \Eprint[0]{\href }%
\@ifxundefined \urlstyle {%
  \providecommand \doi [1]{doi:\discretionary{}{}{}#1}%
}{%
  \providecommand \doi [0]{doi:\discretionary{}{}{}\begingroup
  \urlstyle{rm}\Url }%
}%
\providecommand \doibase [0]{http://dx.doi.org/}%
\providecommand \Doi[1]{\href{\doibase#1}}%
\providecommand \bibAnnote [3]{%
  \BibitemShut{#1}%
  \begin{quotation}\noindent
    \textsc{Key:}\ #2\\\textsc{Annotation:}\ #3%
  \end{quotation}%
}%
\providecommand \bibAnnoteFile [2]{%
  \IfFileExists{#2}{\bibAnnote {#1} {#2} {\input{#2}}}{}%
}%
\providecommand \typeout [0]{\immediate \write \m@ne }%
\providecommand \selectlanguage [0]{\@gobble}%
\providecommand \bibinfo [0]{\@secondoftwo}%
\providecommand \bibfield [0]{\@secondoftwo}%
\providecommand \translation [1]{[#1]}%
\providecommand \BibitemOpen[0]{}%
\providecommand \bibitemStop [0]{}%
\providecommand \bibitemNoStop [0]{.\EOS\space}%
\providecommand \EOS [0]{\spacefactor3000\relax}%
\providecommand \BibitemShut [1]{\csname bibitem#1\endcsname}%
\bibitem{Harris:36}%
  \BibitemOpen
  \bibfield{author}{%
  \bibinfo {author} {\bibfnamefont{S.~E.}\ \bibnamefont{Harris}},\ }%
  \bibfield{title}{%
  \enquote{\bibinfo {title} {Electromagnetically induced transparency},}\ }%
  \bibfield{journal}{%
  \bibinfo {journal} {Phys. Today}\ }%
  \textbf{\bibinfo {volume} {50}},\ \bibinfo {pages} {36--42} (\bibinfo {year}
  {1997})%
  \bibAnnoteFile{NoStop}{Harris:36}%
\bibitem{Fleischhauer:633}%
  \BibitemOpen
  \bibfield{author}{%
  \bibinfo {author} {\bibfnamefont{M.}~\bibnamefont{Fleischhauer}}, \bibinfo
  {author} {\bibfnamefont{A.}~\bibnamefont{Imamo\u{g}lu}},\ and\ \bibinfo
  {author} {\bibfnamefont{J.~P.}\ \bibnamefont{Marangos}},\ }%
  \bibfield{title}{%
  \enquote{\bibinfo {title} {Electromagnetically induced transparency: Optics
  in coherent media},}\ }%
  \bibfield{journal}{%
  \bibinfo {journal} {Rev. Mod. Phys.}\ }%
  \textbf{\bibinfo {volume} {77}},\ \bibinfo {pages} {633--673} (\bibinfo
  {year} {2005})%
  \bibAnnoteFile{NoStop}{Fleischhauer:633}%
\bibitem{Boller:66}%
  \BibitemOpen
  \bibfield{author}{%
  \bibinfo {author} {\bibfnamefont{K.~J.}\ \bibnamefont{Boller}}, \bibinfo
  {author} {\bibfnamefont{A.}~\bibnamefont{Imamo\u{g}lu}},\ and\ \bibinfo
  {author} {\bibfnamefont{S.~E.}\ \bibnamefont{Harris}},\ }%
  \bibfield{title}{%
  \enquote{\bibinfo {title} {Observation of electromagnetically induced
  transparency},}\ }%
  \bibfield{journal}{%
  \bibinfo {journal} {Phys. Rev. Lett.}\ }%
  \textbf{\bibinfo {volume} {66}},\ \bibinfo {pages} {2593--2596} (\bibinfo
  {year} {1991})%
  \bibAnnoteFile{NoStop}{Boller:66}%
\bibitem{Firstenberg:77}%
  \BibitemOpen
  \bibfield{author}{%
  \bibinfo {author} {\bibfnamefont{O.}~\bibnamefont{Firstenberg}}, \bibinfo
  {author} {\bibfnamefont{M.}~\bibnamefont{Shuker}}, \bibinfo {author}
  {\bibfnamefont{R.}~\bibnamefont{Pugatch}}, \bibinfo {author}
  {\bibfnamefont{D.~R.}\ \bibnamefont{Fredkin}}, \bibinfo {author}
  {\bibfnamefont{N.}~\bibnamefont{Davidson}},\ and\ \bibinfo {author}
  {\bibfnamefont{A.}~\bibnamefont{Ron}},\ }%
  \bibfield{title}{%
  \enquote{\bibinfo {title} {Theory of thermal motion in electromagnetically
  induced transparency: Effects of diffusion, {D}oppler broadening, and {D}icke
  and {R}amsey narrowing},}\ }%
  \bibfield{journal}{%
  \bibinfo {journal} {Phys. Rev. A}\ }%
  \textbf{\bibinfo {volume} {77}},\ \bibinfo {pages} {043830} (\bibinfo {year}
  {2008})%
  \bibAnnoteFile{NoStop}{Firstenberg:77}%
\bibitem{Zhang:80}%
  \BibitemOpen
  \bibfield{author}{%
  \bibinfo {author} {\bibfnamefont{Y.}~\bibnamefont{Zhang}}, \bibinfo {author}
  {\bibfnamefont{Z.}~\bibnamefont{Nie}}, \bibinfo {author}
  {\bibfnamefont{H.}~\bibnamefont{Zheng}}, \bibinfo {author}
  {\bibfnamefont{C.}~\bibnamefont{Li}}, \bibinfo {author}
  {\bibfnamefont{J.}~\bibnamefont{Song}},\ and\ \bibinfo {author}
  {\bibfnamefont{M.}~\bibnamefont{Xiao}},\ }%
  \bibfield{title}{%
  \enquote{\bibinfo {title} {Electromagnetically induced spatial nonlinear
  dispersion of four-wave mixing},}\ }%
  \bibfield{journal}{%
  \bibinfo {journal} {Phys. Rev. A}\ }%
  \textbf{\bibinfo {volume} {80}},\ \bibinfo {pages} {013835} (\bibinfo {year}
  {2009})%
  \bibAnnoteFile{NoStop}{Zhang:80}%
\bibitem{Litvak:88}%
  \BibitemOpen
  \bibfield{author}{%
  \bibinfo {author} {\bibfnamefont{A.~G.}\ \bibnamefont{Litvak}}\ and\ \bibinfo
  {author} {\bibfnamefont{M.~D.}\ \bibnamefont{Tokman}},\ }%
  \bibfield{title}{%
  \enquote{\bibinfo {title} {Electromagnetically induced transparency in
  ensembles of classical oscillators},}\ }%
  \bibfield{journal}{%
  \bibinfo {journal} {Phys. Rev. Lett.}\ }%
  \textbf{\bibinfo {volume} {88}},\ \bibinfo {pages} {095003} (\bibinfo {year}
  {2002})%
  \bibAnnoteFile{NoStop}{Litvak:88}%
\bibitem{Shvets:89}%
  \BibitemOpen
  \bibfield{author}{%
  \bibinfo {author} {\bibfnamefont{G.}~\bibnamefont{Shvets}}\ and\ \bibinfo
  {author} {\bibfnamefont{J.~S.}\ \bibnamefont{Wurtele}},\ }%
  \bibfield{title}{%
  \enquote{\bibinfo {title} {Transparency of magnetized plasma at the cyclotron
  frequency},}\ }%
  \bibfield{journal}{%
  \bibinfo {journal} {Phys. Rev. Lett.}\ }%
  \textbf{\bibinfo {volume} {89}},\ \bibinfo {pages} {115003} (\bibinfo {year}
  {2002})%
  \bibAnnoteFile{NoStop}{Shvets:89}%
\bibitem{Werner:61}%
  \BibitemOpen
  \bibfield{author}{%
  \bibinfo {author} {\bibfnamefont{M.~J.}\ \bibnamefont{Werner}}\ and\ \bibinfo
  {author} {\bibfnamefont{A.}~\bibnamefont{Imamo\u{g}lu}},\ }%
  \bibfield{title}{%
  \enquote{\bibinfo {title} {Photon-photon interactions in cavity
  electromagnetically induced transparency},}\ }%
  \bibfield{journal}{%
  \bibinfo {journal} {Phys. Rev. A}\ }%
  \textbf{\bibinfo {volume} {61}},\ \bibinfo {pages} {011801} (\bibinfo {year}
  {1999})%
  \bibAnnoteFile{NoStop}{Werner:61}%
\bibitem{Bentley:61}%
  \BibitemOpen
  \bibfield{author}{%
  \bibinfo {author} {\bibfnamefont{C.~L.}\ \bibnamefont{Bentley}}, \bibinfo
  {author} {\bibfnamefont{J.}~\bibnamefont{Liu}},\ and\ \bibinfo {author}
  {\bibfnamefont{Y.}~\bibnamefont{Liao}},\ }%
  \bibfield{title}{%
  \enquote{\bibinfo {title} {Cavity electromagnetically induced transparency of
  driven-three-level atoms: A transparent window narrowing below a natural
  width},}\ }%
  \bibfield{journal}{%
  \bibinfo {journal} {Phys. Rev. A}\ }%
  \textbf{\bibinfo {volume} {61}},\ \bibinfo {pages} {023811} (\bibinfo {year}
  {2000})%
  \bibAnnoteFile{NoStop}{Bentley:61}%
\bibitem{Dantan:69}%
  \BibitemOpen
  \bibfield{author}{%
  \bibinfo {author} {\bibfnamefont{A.}~\bibnamefont{Dantan}}\ and\ \bibinfo
  {author} {\bibfnamefont{M.}~\bibnamefont{Pinard}},\ }%
  \bibfield{title}{%
  \enquote{\bibinfo {title} {Quantum-state transfer between fields and atoms in
  electromagnetically induced transparency},}\ }%
  \bibfield{journal}{%
  \bibinfo {journal} {Phys. Rev. A}\ }%
  \textbf{\bibinfo {volume} {69}},\ \bibinfo {pages} {043810} (\bibinfo {year}
  {2004})%
  \bibAnnoteFile{NoStop}{Dantan:69}%
\bibitem{Yang:95}%
  \BibitemOpen
  \bibfield{author}{%
  \bibinfo {author} {\bibfnamefont{W.}~\bibnamefont{Yang}}, \bibinfo {author}
  {\bibfnamefont{A.}~\bibnamefont{Joshi}},\ and\ \bibinfo {author}
  {\bibfnamefont{M.}~\bibnamefont{Xiao}},\ }%
  \bibfield{title}{%
  \enquote{\bibinfo {title} {Chaos in an electromagnetically induced
  transparent medium inside an optical cavity},}\ }%
  \bibfield{journal}{%
  \bibinfo {journal} {Phys. Rev. Lett.}\ }%
  \textbf{\bibinfo {volume} {95}},\ \bibinfo {pages} {093902} (\bibinfo {year}
  {2005})%
  \bibAnnoteFile{NoStop}{Yang:95}%
\bibitem{Kuang:76}%
  \BibitemOpen
  \bibfield{author}{%
  \bibinfo {author} {\bibfnamefont{L.~M.}\ \bibnamefont{Kuang}}, \bibinfo
  {author} {\bibfnamefont{Z.~B.}\ \bibnamefont{Chen}},\ and\ \bibinfo {author}
  {\bibfnamefont{J.~W.}\ \bibnamefont{Pan}},\ }%
  \bibfield{title}{%
  \enquote{\bibinfo {title} {Generation of entangled coherent states for
  distant {B}ose-{E}instein condensates via electromagnetically induced
  transparency},}\ }%
  \bibfield{journal}{%
  \bibinfo {journal} {Phys. Rev. A}\ }%
  \textbf{\bibinfo {volume} {76}},\ \bibinfo {pages} {052324} (\bibinfo {year}
  {2007})%
  \bibAnnoteFile{NoStop}{Kuang:76}%
\bibitem{Weatherall:78}%
  \BibitemOpen
  \bibfield{author}{%
  \bibinfo {author} {\bibfnamefont{J.~O.}\ \bibnamefont{Weatherall}}, \bibinfo
  {author} {\bibfnamefont{C.~P.}\ \bibnamefont{Search}},\ and\ \bibinfo
  {author} {\bibfnamefont{M.}~\bibnamefont{J\"a\"askel\"ainen}},\ }%
  \bibfield{title}{%
  \enquote{\bibinfo {title} {Quantum control of electromagnetically induced
  transparency dispersion via atomic tunneling in a double-well
  {B}ose-{E}instein condensate},}\ }%
  \bibfield{journal}{%
  \bibinfo {journal} {Phys. Rev. A}\ }%
  \textbf{\bibinfo {volume} {78}},\ \bibinfo {pages} {013830} (\bibinfo {year}
  {2008})%
  \bibAnnoteFile{NoStop}{Weatherall:78}%
\bibitem{Yamamoto:58}%
  \BibitemOpen
  \bibfield{author}{%
  \bibinfo {author} {\bibfnamefont{K.}~\bibnamefont{Yamamoto}}, \bibinfo
  {author} {\bibfnamefont{K.}~\bibnamefont{Ichimura}},\ and\ \bibinfo {author}
  {\bibfnamefont{N.}~\bibnamefont{Gemma}},\ }%
  \bibfield{title}{%
  \enquote{\bibinfo {title} {Enhanced and reduced absorptions via quantum
  interference: Solid system driven by a rf field},}\ }%
  \bibfield{journal}{%
  \bibinfo {journal} {Phys. Rev. A}\ }%
  \textbf{\bibinfo {volume} {58}},\ \bibinfo {pages} {2460--2466} (\bibinfo
  {year} {1998})%
  \bibAnnoteFile{NoStop}{Yamamoto:58}%
\bibitem{Joshi:67}%
  \BibitemOpen
  \bibfield{author}{%
  \bibinfo {author} {\bibfnamefont{A.}~\bibnamefont{Joshi}}, \bibinfo {author}
  {\bibfnamefont{A.}~\bibnamefont{Brown}}, \bibinfo {author}
  {\bibfnamefont{H.}~\bibnamefont{Wang}},\ and\ \bibinfo {author}
  {\bibfnamefont{M.}~\bibnamefont{Xiao}},\ }%
  \bibfield{title}{%
  \enquote{\bibinfo {title} {Controlling optical bistability in a three-level
  atomic system},}\ }%
  \bibfield{journal}{%
  \bibinfo {journal} {Phys. Rev. A}\ }%
  \textbf{\bibinfo {volume} {67}},\ \bibinfo {pages} {041801} (\bibinfo {year}
  {2003})%
  \bibAnnoteFile{NoStop}{Joshi:67}%
\bibitem{Brown:70}%
  \BibitemOpen
  \bibfield{author}{%
  \bibinfo {author} {\bibfnamefont{A.~W.}\ \bibnamefont{Brown}}\ and\ \bibinfo
  {author} {\bibfnamefont{M.}~\bibnamefont{Xiao}},\ }%
  \bibfield{title}{%
  \enquote{\bibinfo {title} {Modulation transfer in an electromagnetically
  induced transparency system},}\ }%
  \bibfield{journal}{%
  \bibinfo {journal} {Phys. Rev. A}\ }%
  \textbf{\bibinfo {volume} {70}},\ \bibinfo {pages} {053830} (\bibinfo {year}
  {2004})%
  \bibAnnoteFile{NoStop}{Brown:70}%
\bibitem{Yang:72}%
  \BibitemOpen
  \bibfield{author}{%
  \bibinfo {author} {\bibfnamefont{L.}~\bibnamefont{Yang}}, \bibinfo {author}
  {\bibfnamefont{L.}~\bibnamefont{Zhang}}, \bibinfo {author}
  {\bibfnamefont{X.}~\bibnamefont{Li}}, \bibinfo {author}
  {\bibfnamefont{L.}~\bibnamefont{Han}}, \bibinfo {author}
  {\bibfnamefont{G.}~\bibnamefont{Fu}}, \bibinfo {author}
  {\bibfnamefont{N.~B.}\ \bibnamefont{Manson}}, \bibinfo {author}
  {\bibfnamefont{D.}~\bibnamefont{Suter}},\ and\ \bibinfo {author}
  {\bibfnamefont{C.}~\bibnamefont{Wei}},\ }%
  \bibfield{title}{%
  \enquote{\bibinfo {title} {{A}utler-{T}ownes effect in a strongly driven
  electromagnetically induced transparency resonance},}\ }%
  \bibfield{journal}{%
  \bibinfo {journal} {Phys. Rev. A}\ }%
  \textbf{\bibinfo {volume} {72}},\ \bibinfo {pages} {053801} (\bibinfo {year}
  {2005})%
  \bibAnnoteFile{NoStop}{Yang:72}%
\bibitem{Olson:116}%
  \BibitemOpen
  \bibfield{author}{%
  \bibinfo {author} {\bibfnamefont{A.~J.}\ \bibnamefont{Olson}}\ and\ \bibinfo
  {author} {\bibfnamefont{S.~K.}\ \bibnamefont{Mayer}},\ }%
  \bibfield{title}{%
  \enquote{\bibinfo {title} {Electromagnetically induced transparency in
  rubidium},}\ }%
  \bibfield{journal}{%
  \bibinfo {journal} {Am. J. Phys.}\ }%
  \textbf{\bibinfo {volume} {77}},\ \bibinfo {pages} {116--121} (\bibinfo
  {year} {2009})%
  \bibAnnoteFile{NoStop}{Olson:116}%
\bibitem{Joshi:370}%
  \BibitemOpen
  \bibfield{author}{%
  \bibinfo {author} {\bibfnamefont{A.}~\bibnamefont{Joshi}}\ and\ \bibinfo
  {author} {\bibfnamefont{M.}~\bibnamefont{Xiao}},\ }%
  \bibfield{title}{%
  \enquote{\bibinfo {title} {Electromagnetically induced transparency and its
  dispersion properties in a four-level inverted-{Y} atomic system.}.}\ }%
  \bibfield{journal}{%
  \bibinfo {journal} {Phys. Lett. A}\ }%
  \textbf{\bibinfo {volume} {317}},\ \bibinfo {pages} {370} (\bibinfo {year}
  {2003})%
  \bibAnnoteFile{NoStop}{Joshi:370}%
\bibitem{Zhang:99}%
  \BibitemOpen
  \bibfield{author}{%
  \bibinfo {author} {\bibfnamefont{Y.}~\bibnamefont{Zhang}}, \bibinfo {author}
  {\bibfnamefont{A.~W.}\ \bibnamefont{Brown}},\ and\ \bibinfo {author}
  {\bibfnamefont{M.}~\bibnamefont{Xiao}},\ }%
  \bibfield{title}{%
  \enquote{\bibinfo {title} {Opening four-wave mixing and six-wave mixing
  channels via dual electromagnetically induced transparency windows},}\ }%
  \bibfield{journal}{%
  \bibinfo {journal} {Phys. Rev. Lett.}\ }%
  \textbf{\bibinfo {volume} {99}},\ \bibinfo {pages} {123603} (\bibinfo {year}
  {2007})%
  \bibAnnoteFile{NoStop}{Zhang:99}%
\bibitem{Li:101}%
  \BibitemOpen
  \bibfield{author}{%
  \bibinfo {author} {\bibfnamefont{S.}~\bibnamefont{Li}}, \bibinfo {author}
  {\bibfnamefont{X.}~\bibnamefont{Yang}}, \bibinfo {author}
  {\bibfnamefont{X.}~\bibnamefont{Cao}}, \bibinfo {author}
  {\bibfnamefont{C.}~\bibnamefont{Zhang}}, \bibinfo {author}
  {\bibfnamefont{C.}~\bibnamefont{Xie}},\ and\ \bibinfo {author}
  {\bibfnamefont{H.}~\bibnamefont{Wang}},\ }%
  \bibfield{title}{%
  \enquote{\bibinfo {title} {Enhanced cross-phase modulation based on a double
  electromagnetically induced transparency in a four-level tripod atomic
  system},}\ }%
  \bibfield{journal}{%
  \bibinfo {journal} {Phys. Rev. Lett.}\ }%
  \textbf{\bibinfo {volume} {101}},\ \bibinfo {pages} {073602} (\bibinfo {year}
  {2008})%
  \bibAnnoteFile{NoStop}{Li:101}%
\bibitem{Joshi:79}%
  \BibitemOpen
  \bibfield{author}{%
  \bibinfo {author} {\bibfnamefont{A.}~\bibnamefont{Joshi}},\ }%
  \bibfield{title}{%
  \enquote{\bibinfo {title} {Phase-dependent electromagnetically induced
  transparency and its dispersion properties in a four-level quantum well
  system},}\ }%
  \bibfield{journal}{%
  \bibinfo {journal} {Phys. Rev. B}\ }%
  \textbf{\bibinfo {volume} {79}},\ \bibinfo {pages} {115315} (\bibinfo {year}
  {2009})%
  \bibAnnoteFile{NoStop}{Joshi:79}%
\bibitem{Xiao:666}%
  \BibitemOpen
  \bibfield{author}{%
  \bibinfo {author} {\bibfnamefont{M.}~\bibnamefont{Xiao}}, \bibinfo {author}
  {\bibfnamefont{Y.~Q.}\ \bibnamefont{Li}}, \bibinfo {author}
  {\bibfnamefont{S.~Z.}\ \bibnamefont{Jin}},\ and\ \bibinfo {author}
  {\bibfnamefont{J.}~\bibnamefont{Gea-Banacloche}},\ }%
  \bibfield{title}{%
  \enquote{\bibinfo {title} {Measurement of dispersive properties of
  electromagnetically induced transparency in rubidium atoms},}\ }%
  \bibfield{journal}{%
  \bibinfo {journal} {Phys. Rev. Lett.}\ }%
  \textbf{\bibinfo {volume} {74}},\ \bibinfo {pages} {666--669} (\bibinfo
  {year} {1995})%
  \bibAnnoteFile{NoStop}{Xiao:666}%
\bibitem{Hau:397}%
  \BibitemOpen
  \bibfield{author}{%
  \bibinfo {author} {\bibfnamefont{L.~V.}\ \bibnamefont{Hau}}, \bibinfo
  {author} {\bibfnamefont{S.~E.}\ \bibnamefont{Harris}}, \bibinfo {author}
  {\bibfnamefont{Z.}~\bibnamefont{Dutton}},\ and\ \bibinfo {author}
  {\bibfnamefont{C.~H.}\ \bibnamefont{Behroozi}},\ }%
  \bibfield{title}{%
  \enquote{\bibinfo {title} {Light speed reduction to 17 m/s in an ultracold
  atomic gas},}\ }%
  \bibfield{journal}{%
  \bibinfo {journal} {Nature}\ }%
  \textbf{\bibinfo {volume} {397}},\ \bibinfo {pages} {594--598} (\bibinfo
  {year} {1999})%
  \bibAnnoteFile{NoStop}{Hau:397}%
\bibitem{Liu:409}%
  \BibitemOpen
  \bibfield{author}{%
  \bibinfo {author} {\bibfnamefont{C.}~\bibnamefont{Liu}}, \bibinfo {author}
  {\bibfnamefont{Z.}~\bibnamefont{Dutton}}, \bibinfo {author}
  {\bibfnamefont{C.~H.}\ \bibnamefont{Behroozi}},\ and\ \bibinfo {author}
  {\bibfnamefont{L.~V.}\ \bibnamefont{Hau}},\ }%
  \bibfield{title}{%
  \enquote{\bibinfo {title} {Observation of coherent optical information
  storage in an atomic medium using halted light pulses},}\ }%
  \bibfield{journal}{%
  \bibinfo {journal} {Nature}\ }%
  \textbf{\bibinfo {volume} {409}},\ \bibinfo {pages} {490--493} (\bibinfo
  {year} {2001})%
  \bibAnnoteFile{NoStop}{Liu:409}%
\bibitem{Mompart:R7}%
  \BibitemOpen
  \bibfield{author}{%
  \bibinfo {author} {\bibfnamefont{J.}~\bibnamefont{Mompart}}\ and\ \bibinfo
  {author} {\bibfnamefont{R.}~\bibnamefont{Corbal\'{a}n}},\ }%
  \bibfield{title}{%
  \enquote{\bibinfo {title} {Lasing without inversion},}\ }%
  \bibfield{journal}{%
  \bibinfo {journal} {J. Opt. B: Quantum Semiclassical Opt.}\ }%
  \textbf{\bibinfo {volume} {2}},\ \bibinfo {pages} {R7} (\bibinfo {year}
  {2000})%
  \bibAnnoteFile{NoStop}{Mompart:R7}%
\bibitem{Wu:78}%
  \BibitemOpen
  \bibfield{author}{%
  \bibinfo {author} {\bibfnamefont{H.}~\bibnamefont{Wu}}, \bibinfo {author}
  {\bibfnamefont{M.}~\bibnamefont{Xiao}},\ and\ \bibinfo {author}
  {\bibfnamefont{J.}~\bibnamefont{Gea-Banacloche}},\ }%
  \bibfield{title}{%
  \enquote{\bibinfo {title} {Evidence of lasing without inversion in a hot
  rubidium vapor under electromagnetically-induced-transparency conditions},}\
  }%
  \bibfield{journal}{%
  \bibinfo {journal} {Phys. Rev. A}\ }%
  \textbf{\bibinfo {volume} {78}},\ \bibinfo {pages} {041802} (\bibinfo {year}
  {2008})%
  \bibAnnoteFile{NoStop}{Wu:78}%
\bibitem{Wang:87}%
  \BibitemOpen
  \bibfield{author}{%
  \bibinfo {author} {\bibfnamefont{H.}~\bibnamefont{Wang}}, \bibinfo {author}
  {\bibfnamefont{D.}~\bibnamefont{Goorskey}},\ and\ \bibinfo {author}
  {\bibfnamefont{M.}~\bibnamefont{Xiao}},\ }%
  \bibfield{title}{%
  \enquote{\bibinfo {title} {Enhanced {K}err nonlinearity via atomic coherence
  in a three-level atomic system},}\ }%
  \bibfield{journal}{%
  \bibinfo {journal} {Phys. Rev. Lett.}\ }%
  \textbf{\bibinfo {volume} {87}},\ \bibinfo {pages} {073601} (\bibinfo {year}
  {2001})%
  \bibAnnoteFile{NoStop}{Wang:87}%
\bibitem{Schmidt:76}%
  \BibitemOpen
  \bibfield{author}{%
  \bibinfo {author} {\bibfnamefont{H.}~\bibnamefont{Schmidt}}\ and\ \bibinfo
  {author} {\bibfnamefont{R.~J.}\ \bibnamefont{Ram}},\ }%
  \bibfield{title}{%
  \enquote{\bibinfo {title} {All-optical wavelength converter and switch based
  on electromagnetically induced transparency},}\ }%
  \bibfield{journal}{%
  \bibinfo {journal} {Appl. Phys. Lett.}\ }%
  \textbf{\bibinfo {volume} {76}},\ \bibinfo {pages} {3173--3175} (\bibinfo
  {year} {2000})%
  \bibAnnoteFile{NoStop}{Schmidt:76}%
\bibitem{Ottaviani:90}%
  \BibitemOpen
  \bibfield{author}{%
  \bibinfo {author} {\bibfnamefont{C.}~\bibnamefont{Ottaviani}}, \bibinfo
  {author} {\bibfnamefont{D.}~\bibnamefont{Vitali}}, \bibinfo {author}
  {\bibfnamefont{M.}~\bibnamefont{Artoni}}, \bibinfo {author}
  {\bibfnamefont{F.}~\bibnamefont{Cataliotti}},\ and\ \bibinfo {author}
  {\bibfnamefont{P.}~\bibnamefont{Tombesi}},\ }%
  \bibfield{title}{%
  \enquote{\bibinfo {title} {Polarization qubit phase gate in driven atomic
  media},}\ }%
  \bibfield{journal}{%
  \bibinfo {journal} {Phys. Rev. Lett.}\ }%
  \textbf{\bibinfo {volume} {90}},\ \bibinfo {pages} {197902} (\bibinfo {year}
  {2003})%
  \bibAnnoteFile{NoStop}{Ottaviani:90}%
\bibitem{Hetet:77}%
  \BibitemOpen
  \bibfield{author}{%
  \bibinfo {author} {\bibfnamefont{G.}~\bibnamefont{H\'etet}}, \bibinfo
  {author} {\bibfnamefont{A.}~\bibnamefont{Peng}}, \bibinfo {author}
  {\bibfnamefont{M.~T.}\ \bibnamefont{Johnsson}}, \bibinfo {author}
  {\bibfnamefont{J.~J.}\ \bibnamefont{Hope}},\ and\ \bibinfo {author}
  {\bibfnamefont{P.~K.}\ \bibnamefont{Lam}},\ }%
  \bibfield{title}{%
  \enquote{\bibinfo {title} {Characterization of
  electromagnetically-induced-transparency-based continuous-variable quantum
  memories},}\ }%
  \bibfield{journal}{%
  \bibinfo {journal} {Phys. Rev. A}\ }%
  \textbf{\bibinfo {volume} {77}},\ \bibinfo {pages} {012323} (\bibinfo {year}
  {2008})%
  \bibAnnoteFile{NoStop}{Hetet:77}%
\bibitem{Bermel:74}%
  \BibitemOpen
  \bibfield{author}{%
  \bibinfo {author} {\bibfnamefont{P.}~\bibnamefont{Bermel}}, \bibinfo {author}
  {\bibfnamefont{A.}~\bibnamefont{Rodr\'{\i}guez}}, \bibinfo {author}
  {\bibfnamefont{S.~G.}\ \bibnamefont{Johnson}}, \bibinfo {author}
  {\bibfnamefont{J.~D.}\ \bibnamefont{Joannopoulos}},\ and\ \bibinfo {author}
  {\bibfnamefont{M.}~\bibnamefont{Solja\v{c}i\'{c}}},\ }%
  \bibfield{title}{%
  \enquote{\bibinfo {title} {Single-photon all-optical switching using
  waveguide-cavity quantum electrodynamics},}\ }%
  \bibfield{journal}{%
  \bibinfo {journal} {Phys. Rev. A}\ }%
  \textbf{\bibinfo {volume} {74}},\ \bibinfo {pages} {043818} (\bibinfo {year}
  {2006})%
  \bibAnnoteFile{NoStop}{Bermel:74}%
\bibitem{Dragoman}%
  \BibitemOpen
  \bibfield{author}{%
  \bibinfo {author} {\bibfnamefont{D.}~\bibnamefont{Dragoman}}\ and\ \bibinfo
  {author} {\bibfnamefont{M.}~\bibnamefont{Dragoman}},\ }%
  \emph{\bibinfo {title} {Quantum-Classical Analogies}}\ (\bibinfo {publisher}
  {Springer},\ \bibinfo {address} {Berlin},\ \bibinfo {year} {2004})%
  \bibAnnoteFile{NoStop}{Dragoman}%
\bibitem{Hemmer:88}%
  \BibitemOpen
  \bibfield{author}{%
  \bibinfo {author} {\bibfnamefont{P.~R.}\ \bibnamefont{Hemmer}}\ and\ \bibinfo
  {author} {\bibfnamefont{M.~G.}\ \bibnamefont{Prentiss}},\ }%
  \bibfield{title}{%
  \enquote{\bibinfo {title} {Coupled-pendulum model of the stimulated resonance
  {R}aman effect},}\ }%
  \bibfield{journal}{%
  \bibinfo {journal} {J. Opt. Soc. Am. B}\ }%
  \textbf{\bibinfo {volume} {5}},\ \bibinfo {pages} {1613--1623} (\bibinfo
  {year} {1988})%
  \bibAnnoteFile{NoStop}{Hemmer:88}%
\bibitem{Shore:77}%
  \BibitemOpen
  \bibfield{author}{%
  \bibinfo {author} {\bibfnamefont{B.~W.}\ \bibnamefont{Shore}}, \bibinfo
  {author} {\bibfnamefont{M.~V.}\ \bibnamefont{Gromovyy}}, \bibinfo {author}
  {\bibfnamefont{L.~P.}\ \bibnamefont{Yatsenko}},\ and\ \bibinfo {author}
  {\bibfnamefont{V.~I.}\ \bibnamefont{Romanenko}},\ }%
  \bibfield{title}{%
  \enquote{\bibinfo {title} {Simple mechanical analogs of rapid adiabatic
  passage in atomic physics},}\ }%
  \bibfield{journal}{%
  \bibinfo {journal} {Am. J. Phys.}\ }%
  \textbf{\bibinfo {volume} {77}},\ \bibinfo {pages} {1183--1194} (\bibinfo
  {year} {2009})%
  \bibAnnoteFile{NoStop}{Shore:77}%
\bibitem{Zhu:64}%
  \BibitemOpen
  \bibfield{author}{%
  \bibinfo {author} {\bibfnamefont{Y.}~\bibnamefont{Zhu}}, \bibinfo {author}
  {\bibfnamefont{D.~J.}\ \bibnamefont{Gauthier}}, \bibinfo {author}
  {\bibfnamefont{S.~E.}\ \bibnamefont{Morin}}, \bibinfo {author}
  {\bibfnamefont{Q.}~\bibnamefont{Wu}}, \bibinfo {author}
  {\bibfnamefont{H.~J.}\ \bibnamefont{Carmichael}},\ and\ \bibinfo {author}
  {\bibfnamefont{T.~W.}\ \bibnamefont{Mossberg}},\ }%
  \bibfield{title}{%
  \enquote{\bibinfo {title} {Vacuum rabi splitting as a feature of
  linear-dispersion theory: Analysis and experimental observations},}\ }%
  \bibfield{journal}{%
  \bibinfo {journal} {Phys. Rev. Lett.}\ }%
  \textbf{\bibinfo {volume} {64}},\ \bibinfo {pages} {2499--2502} (\bibinfo
  {year} {1990})%
  \bibAnnoteFile{NoStop}{Zhu:64}%
\bibitem{Vaccaro:243}%
  \BibitemOpen
  \bibfield{author}{%
  \bibinfo {author} {\bibfnamefont{J.~A.}\ \bibnamefont{Vaccaro}}\ and\
  \bibinfo {author} {\bibfnamefont{A.}~\bibnamefont{Joshi}},\ }%
  \bibfield{title}{%
  \enquote{\bibinfo {title} {Position-momentum and number-phase wigner
  functions and their respective displacement operators},}\ }%
  \bibfield{journal}{%
  \bibinfo {journal} {Phys. Lett. A}\ }%
  \textbf{\bibinfo {volume} {243}},\ \bibinfo {pages} {13--19} (\bibinfo {year}
  {1998})%
  \bibAnnoteFile{NoStop}{Vaccaro:243}%
\bibitem{Garrido:37}%
  \BibitemOpen
  \bibfield{author}{%
  \bibinfo {author} {\bibfnamefont{C.~L.}\ \bibnamefont{{Garrido Alzar}}},
  \bibinfo {author} {\bibfnamefont{M.~A.~G.}\ \bibnamefont{Mart\'{\i}nez}},\
  and\ \bibinfo {author} {\bibfnamefont{P.}~\bibnamefont{Nussenzveig}},\ }%
  \bibfield{title}{%
  \enquote{\bibinfo {title} {Classical analog of electromagnetically induced
  transparency},}\ }%
  \bibfield{journal}{%
  \bibinfo {journal} {Am. J. Phys.}\ }%
  \textbf{\bibinfo {volume} {70}},\ \bibinfo {pages} {37--41} (\bibinfo {year}
  {2002})%
  \bibAnnoteFile{NoStop}{Garrido:37}%
\bibitem{Lifshitz:67}%
  \BibitemOpen
  \bibfield{author}{%
  \bibinfo {author} {\bibfnamefont{R.}~\bibnamefont{Lifshitz}}\ and\ \bibinfo
  {author} {\bibfnamefont{M.~C.}\ \bibnamefont{Cross}},\ }%
  \bibfield{title}{%
  \enquote{\bibinfo {title} {Response of parametrically driven nonlinear
  coupled oscillators with application to micromechanical and nanomechanical
  resonator arrays},}\ }%
  \bibfield{journal}{%
  \bibinfo {journal} {Phys. Rev. B}\ }%
  \textbf{\bibinfo {volume} {67}},\ \bibinfo {pages} {134302} (\bibinfo {year}
  {2003})%
  \bibAnnoteFile{NoStop}{Lifshitz:67}%
\bibitem{Allen-Eberly}%
  \BibitemOpen
  \bibfield{author}{%
  \bibinfo {author} {\bibfnamefont{L.}~\bibnamefont{Allen}}\ and\ \bibinfo
  {author} {\bibfnamefont{J.~H.}\ \bibnamefont{Eberly}},\ }%
  \emph{\bibinfo {title} {Optical Resonance and Two-Level Atoms}}\ (\bibinfo
  {publisher} {Dover},\ \bibinfo {address} {New York},\ \bibinfo {year}
  {1987})%
  \bibAnnoteFile{NoStop}{Allen-Eberly}%
\bibitem{Petrosyan:65}%
  \BibitemOpen
  \bibfield{author}{%
  \bibinfo {author} {\bibfnamefont{D.}~\bibnamefont{Petrosyan}}\ and\ \bibinfo
  {author} {\bibfnamefont{G.}~\bibnamefont{Kurizki}},\ }%
  \bibfield{title}{%
  \enquote{\bibinfo {title} {Symmetric photon-photon coupling by atoms with
  {Z}eeman-split sublevels},}\ }%
  \bibfield{journal}{%
  \bibinfo {journal} {Phys. Rev. A}\ }%
  \textbf{\bibinfo {volume} {65}},\ \bibinfo {pages} {033833} (\bibinfo {year}
  {2002})%
  \bibAnnoteFile{NoStop}{Petrosyan:65}%
\bibitem{Autler:703}%
  \BibitemOpen
  \bibfield{author}{%
  \bibinfo {author} {\bibfnamefont{S.~H.}\ \bibnamefont{Autler}}\ and\ \bibinfo
  {author} {\bibfnamefont{C.~H.}\ \bibnamefont{Townes}},\ }%
  \bibfield{title}{%
  \enquote{\bibinfo {title} {Stark effect in rapidly varying fields},}\ }%
  \bibfield{journal}{%
  \bibinfo {journal} {Phys. Rev.}\ }%
  \textbf{\bibinfo {volume} {100}},\ \bibinfo {pages} {703--722} (\bibinfo
  {year} {1955})%
  \bibAnnoteFile{NoStop}{Autler:703}%
\bibitem{Li:43}%
  \BibitemOpen
  \bibfield{author}{%
  \bibinfo {author} {\bibfnamefont{Yong-qing}\ \bibnamefont{Li}}\ and\ \bibinfo
  {author} {\bibfnamefont{Min}\ \bibnamefont{Xiao}},\ }%
  \bibfield{title}{%
  \enquote{\bibinfo {title} {Electromagnetically induced transparency in a
  three-level $\lambda$-type system in rubidium atoms},}\ }%
  \bibfield{journal}{%
  \bibinfo {journal} {Phys. Rev. A}\ }%
  \textbf{\bibinfo {volume} {51}},\ \bibinfo {pages} {R2703--R2706} (\bibinfo
  {year} {1995})%
  \bibAnnoteFile{NoStop}{Li:43}%
\bibitem{Li-S:44}%
  \BibitemOpen
  \bibfield{author}{%
  \bibinfo {author} {\bibfnamefont{S.}~\bibnamefont{Li}}, \bibinfo {author}
  {\bibfnamefont{X.}~\bibnamefont{Yang}}, \bibinfo {author}
  {\bibfnamefont{X.}~\bibnamefont{Cao}}, \bibinfo {author}
  {\bibfnamefont{C.}~\bibnamefont{Xie}},\ and\ \bibinfo {author}
  {\bibfnamefont{H.}~\bibnamefont{Wang}},\ }%
  \bibfield{title}{%
  \enquote{\bibinfo {title} {Two electromagnetically induced transparency
  windows and an enhanced electromagnetically induced transparency signal in a
  four-level tripod atomic system},}\ }%
  \bibfield{journal}{%
  \bibinfo {journal} {J. Phys. B: At. Mol. Opt. Phys.}\ }%
  \textbf{\bibinfo {volume} {40}},\ \bibinfo {pages} {3211} (\bibinfo {year}
  {2007})%
  \bibAnnoteFile{NoStop}{Li-S:44}%
\bibitem{Lorentz}%
  \BibitemOpen
  \bibfield{author}{%
  \bibinfo {author} {\bibfnamefont{H.~A.}\ \bibnamefont{Lorentz}},\ }%
  \emph{\bibinfo {title} {The Theory of Electrons}}\ (\bibinfo {publisher}
  {Dover},\ \bibinfo {address} {New York},\ \bibinfo {year} {1952})\
  Chap.~\bibinfo {chapter} {4}%
  \bibAnnoteFile{NoStop}{Lorentz}%
\bibitem{French}%
  \BibitemOpen
  \bibfield{author}{%
  \bibinfo {author} {\bibfnamefont{A.~P.}\ \bibnamefont{French}},\ }%
  \emph{\bibinfo {title} {Vibrations and Waves}}\ (\bibinfo {publisher}
  {Norton},\ \bibinfo {address} {New York},\ \bibinfo {year} {1971})\ pp.\
  \bibinfo {pages} {96--101}%
  \bibAnnoteFile{NoStop}{French}%
\bibitem{Scully-Zubairy}%
  \BibitemOpen
  \bibfield{author}{%
  \bibinfo {author} {\bibfnamefont{M.~O.}\ \bibnamefont{Scully}}\ and\ \bibinfo
  {author} {\bibfnamefont{M.~S.}\ \bibnamefont{Zubairy}},\ }%
  \emph{\bibinfo {title} {Quantum Optics}}\ (\bibinfo {publisher} {Cambridge
  University Press},\ \bibinfo {address} {Cambridge, New York},\ \bibinfo
  {year} {1997})\ pp.\ \bibinfo {pages} {2--9}%
  \bibAnnoteFile{NoStop}{Scully-Zubairy}%
\bibitem{Marion-Thornton}%
  \BibitemOpen
  \bibfield{author}{%
  \bibinfo {author} {\bibfnamefont{J.~B.}\ \bibnamefont{Marion}}\ and\ \bibinfo
  {author} {\bibfnamefont{S.~T.}\ \bibnamefont{Thornton}},\ }%
  \emph{\bibinfo {title} {Classical Dynamics of Particles and Systems}},\
  \bibinfo {edition} {4th}\ ed.\ (\bibinfo {publisher} {Saunders},\ \bibinfo
  {address} {Fort Worth, Philadelphia},\ \bibinfo {year} {1995})\ pp.\ \bibinfo
  {pages} {131--137}%
  \bibAnnoteFile{NoStop}{Marion-Thornton}%
\bibitem{Symon}%
  \BibitemOpen
  \bibfield{author}{%
  \bibinfo {author} {\bibfnamefont{K.~R.}\ \bibnamefont{Symon}},\ }%
  \emph{\bibinfo {title} {Mechanics}},\ \bibinfo {edition} {3rd}\ ed.\
  (\bibinfo {publisher} {Addison-Wesley},\ \bibinfo {address} {Reading,
  Mass.},\ \bibinfo {year} {1971})\ p.\ \bibinfo {pages} {201}%
  \bibAnnoteFile{NoStop}{Symon}%
\end{thebibliography}%

\end{document}